\documentclass[twocolumn,aps,pra,showpacs,superscriptaddress,notitlepage,9pt]{revtex4-1}

\usepackage{amsmath}
\usepackage{amsfonts}
\usepackage{amssymb}
\usepackage{graphicx}
\usepackage{braket}
\newcommand{\dg}{^{\dagger}}
\usepackage{lipsum}
\usepackage{fullpage}
\usepackage{appendix}
\usepackage{color}

\newcommand{\hc}{\text{H.c.}}
\newcommand{\eq}[1]{\begin{align}#1\end{align}}
\newcommand{\seq}[1]{\begin{subequations}#1\end{subequations}}

\begin{document}
\title{Non-Hermitian engineering for brighter broadband  pseudothermal light}
\author{Nicol\'as Quesada}
\affiliation{Perimeter Institute for Theoretical Physics, Waterloo, Ontario, N2L 2Y5, Canada}
\author{Eugene Adjei}
\affiliation{Perimeter Institute for Theoretical Physics, Waterloo, Ontario, N2L 2Y5, Canada}
\affiliation{Department of Physics \& Astronomy, University of Waterloo, Waterloo, Ontario, Canada, N2L 3G1}
\author{Ramy El-Ganainy}
\affiliation{Department of Physics, Department of Electrical and Computer Engineering,  and Henes Center for Quantum Phenomena, Michigan Technological University, Houghton, Michigan, 49931, USA}
\author{Agata M. Bra\'nczyk}
\affiliation{Perimeter Institute for Theoretical Physics, Waterloo, Ontario, N2L 2Y5, Canada}

\begin{abstract}
We show that non-Hermitian engineering can play a positive role  in quantum systems.  This is in contrast to the widely accepted notion  that optical losses are a foe that must be eliminated or, at least, minimized. We take advantage of the interplay between nonlinear interactions and loss to show that spectral-loss engineering can relax phase-matching conditions, enabling generation of broadband pseudothermal states at new frequencies. This opens the door for utilizing the full potential of semiconductor materials that exhibit giant nonlinearities but lack the necessary ingredients for achieving quasi-phase matching. This in turn may pave the way for building on-chip quantum light sources.
\end{abstract}

\maketitle

\section{Introduction} 
Recent developments in parity-time (PT) symmetric, and general non-Hermitian, optics \cite{El-Ganainy2007OL,El-Ganainy2008PRL-B,El-Ganainy2008PRL-O,El-Ganainy2010NP,Hodaei2014S,Feng2014S,Peng2014NP,Hodaei2017N,Chen2017N} continue to generate intriguing results at both the fundamental and engineering levels. Importantly, these works are changing the widely-accepted notion that optical losses are a foe that must be eliminated or, at least, minimized. On the contrary,  engineering the interplay between loss and gain (or neutral elements) was recently shown to lead to unexpected effects such as loss-induced lasing, laser self-termination \cite{Liertzer2012PRL,Brandstetter2014NC,Peng2014S,El-Ganainy2014PRA} and unidirectional invisibility \cite{Lin2011PRL,Zhu2013OL,Longhi2015OL}, to just mention a few examples. For recent reviews, see \cite{Feng2017NP,El-Ganainy2018NP}. 

Given this intense activity, it is perhaps surprising that the exploration of non-Hermitian engineering in the quantum regime has been relatively limited in scope, mainly emphasizing the limitations imposed by quantum noise on non-Hermitian systems \cite{Schomerus2010, Agarwal2012,Min2015,Makris2016,Vashahri-Ghamsari2017}. In addition, it is also well-established that losses (and other forms of coupling to the environment) in quantum systems is a main source of decoherence \cite{Schlosshauer2005}. One can thus wonder if optical losses can be useful at all in quantum engineering.

In this paper, we address this question in the context of spontaneous generation of broadband pseudothermal states in one output mode of a wave-mixing process \cite{Christ2011,Spasibko2017,Vernon2017,khandekar2017near}. Spontaneous photon-generation is an inherently quantum-mechanical process, and although thermal states are diagonal in the photon-number and coherent-state bases, they can behave non-classically \cite{Kim2002a,Zavatta2007,Tahira2009,Weedbrook2012,Harder2014,Luis2017,deng2019quantum}. 

\begin{figure}[b!]
	\begin{center}
		\includegraphics[width=\columnwidth]{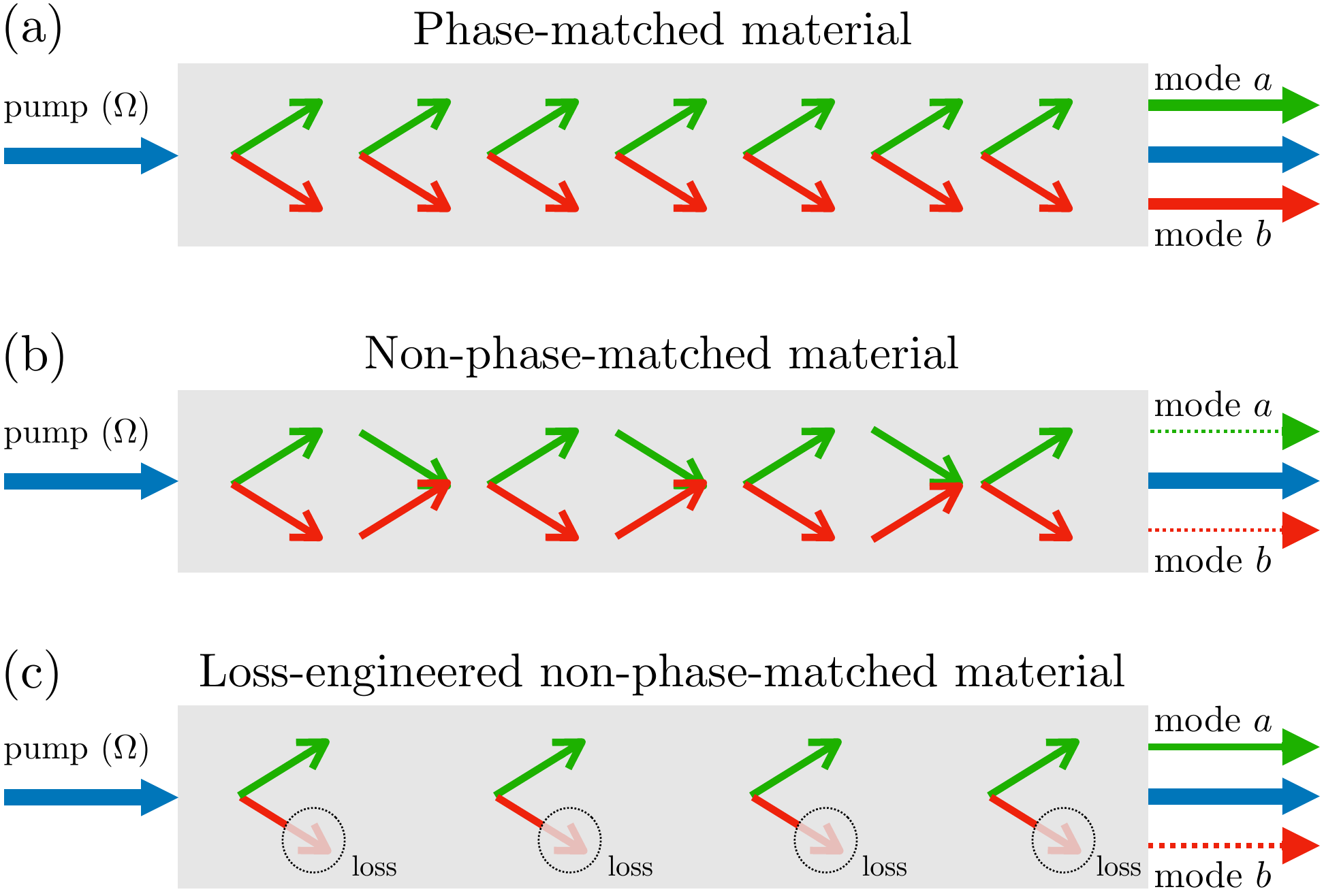}
	\end{center}
	\vspace{-0.3cm}
	\caption{ (a) A phase-matched material facilitates pair creation (internal arrows pointing out), generating intense output in modes $a$ and $b$. (b)  Away from phase-matching, oscillations in the fields' relative phases  spawn two competing processes: pair creation (outward arrows) and  recombination (inward arrows), generating negligible output in modes $a$ and $b$. (c) Adding loss in mode $a$ of a non-phase-matched material suppresses recombination without disrupting pair creation, generating  enhanced output in mode $b$.}
	\label{recomb}
\end{figure}

Our proposed scheme relies on the interplay between nonlinear interactions and loss (see Figure \ref{recomb}).  A pump beam is incident on a nonlinear medium with second- or third-order nonlinearity. Pump photons can then spontaneously convert into a pair of signal and idler photons. When the system is Hermitian, stringent phase-matching conditions must be satisfied in order for this conversion to be efficient, which poses serious limitations on building on-chip quantum-light-sources based on wave-mixing processes. It is conceivable, however, that by introducing optical losses to the idler component, one can force an efficient irreversible down (up) conversion, while at the same time relaxing the phase-matching condition.

The intuition for this effect is as follows. In a \emph{phase-matched material}, the non-linear interaction facilitates pair creation in modes $a$ and $b$, generating intense output in these modes (Fig. \ref{recomb} (a)). But \emph{away from phase-matching}, oscillations in the fields' relative phases  spawn two competing processes: pair creation and  recombination (Fig. \ref{recomb} (b)). These competing processes interfere, and thus  negligible output is generated in modes $a$ and $b$. To first order, this undesirable interference effect can be prevented by adding loss in, e.g., mode $b$ of the non-phase-matched material, which suppresses recombination (because if a photon in mode $b$ is lost, it cannot recombine with the photon in mode $a$) without disrupting pair creation, which leads to enhanced output in mode $a$ (Fig. \ref{recomb} (c)). More rigorously, one must consider the full dynamics of the system, which reveals that increasing the losses beyond a certain threshold value can indeed impede the generation process. Thus, as we will show shortly, increasing the idler losses leads to a competition between the generation and recombination processes and the ideal behavior occur at an optimal loss value.

While these ideas were recently proposed for building classical on-chip nonlinear light sources \cite{El-Ganainy2015OL,Zhong2016NJP} and can be traced back to loss-induced modulation instability in nonlinear fiber optics \cite{Perego2018}, it is not a priori clear if similar concepts can be applied successfully in the quantum regime due to the effect of quantum noise. Here we demonstrate that even when quantum fluctuations are relevant, non-Hermiticity can still play a positive role. 

\section{Formalism} 

We start by considering the process of twin-beam generation, into modes $a$ and $b$, due to spontaneous three- or four-wave mixing  (assuming a cw pump at frequency $\Omega$).

The evolution of the beams inside the nonlinear material is governed by the momentum operator (see Appendix \ref{sec:H}, which follows the treatment in Refs. \cite{wasilewski2006pulsed,caves1987quantum,shirasaki1990squeezing}):
\begin{align}
P    ={}& \hbar \int d\omega [\Delta k_a (\omega) a\dg(\omega) a(\omega)+\Delta k_b (\omega) b\dg(\omega) b(\omega)] \nonumber\\ 
& + \hbar \xi g(z) \int d\omega a\dg(\omega)b\dg(\Omega-\omega)+\text{h.c.} 
\end{align}
The  operator $P$ generates translation along the longitudinal axis $z$ of the nonlinear medium, in analogy to how a Hamiltonian generates translation in time. The parameter $\xi$, which depends on the peak material nonlinearity and the peak pump amplitude, determines the strength of the interaction. The field operators $a(\omega)$ and $b(\omega)$ annihilate photons at frequency $\omega$ in modes $a$ and $b$ respectively, and satisfy the commutation relations $[a(\omega),a\dg(\omega')]=[b(\omega),b\dg(\omega')]=\delta(\omega-\omega')$. The functions $\Delta k_j(\omega)$ determine the phase-matching inside the material (which  amounts to satisfying momentum conservation) and are defined in Appendix \ref{sec:H}.  Finally, the function $g(z)$ is the normalized nonlinearity profile of the material (which can be customized using  nonlinearity shaping methods \cite{Branczyk2011,Dixon2013,Dosseva2016,Tambasco2016,Graffitti2017}). Here, we take  $g(z)$ to be a rectangular function $\Pi_{0,L}(z)=1$ for $0<z<L$ and $\Pi_{0,L}(z)=0$ otherwise. 

In the absence of optical loss, the state generated by the operator $P$ will be a spectrally entangled twin-beam state, i.e., a manifold of two-mode squeezed vacua (see Appendix \ref{BPT}).  Despite a cw pump, the spectrum of each individual beam will be broad, as opposed to cw, with a bandwidth that gets more narrow as the length of the nonlinear region increases \cite{Quesada2019}. The reduced state of each beam will thus be a broadband pseudothermal state.

We now assume that all the frequencies of mode $b$ experience the same optical loss at a rate $\gamma_{b}$. This can be modeled using the Lindblad master equation:
\begin{align} \label{Lindblad}
\begin{split}
\frac{\partial}{\partial z}\rho={}&-\frac{i}{\hbar}[\rho,P]+ \gamma_b\int d\omega \Big(b(\omega)\rho b\dg(\omega)\\
&-\frac{1}{2}\{b\dg(\omega) b(\omega),\rho\}\Big)\,.
\end{split}
\end{align}
Rather than solving for $\rho$ directly, we use  expressions for the spectral densities $n_a$ and $n_b$ in modes $a$ and $b$, as well as the cross correlation $m$ between the two modes: 
\begin{subequations} \label{eq:n}
\begin{align}
&\braket{a\dg(\omega)a(\omega')} = n_a(\omega) \delta(\omega-\omega')\\
&\braket{b\dg(\Omega-\omega)b(\Omega-\omega')} = n_b(\omega) \delta(\omega-\omega')\\ 
&\langle a(\omega) b(\Omega-\omega') \rangle = m(\omega) \delta(\omega-\omega'),
\end{align}
\end{subequations}
to obtain (see Appendix \ref{app:eoms}): 
\begin{subequations} \label{eq:dndz}
\begin{align} 
\frac{dn_a(\omega)}{dz} ={}& \text{i}\xi m(\omega) - \text{i}\xi m^*(\omega)\,\\
\frac{dn_b(\omega)}{dz} = {}&\text{i}\xi m(\omega) - \text{i}\xi m^*(\omega) - \gamma_{b}n_b(\omega)\,\\
\begin{split}
\frac{dm(\omega)}{dz} ={}& -\text{i}\Delta k(\omega) m(\omega)-\frac{\gamma_{b}}{2}m(\omega)\\\label{eq:dmdz}
& -\text{i}\xi n_a(\omega) - \text{i}\xi n_b(\omega)-\text{i}\xi\,,
\end{split}
\end{align}
\end{subequations}
where $\Delta k(\omega) = \Delta k_{a}(\omega) + \Delta k_{b}(\omega)$. Equations  (\ref{eq:n}) reveal that, despite having a broadband output, the frequencies within each of the output modes are completely decorrelated. This is a consequence of the fact that we assumed a cw, quasi-monochromatic, pump that enforces strict energy conservation $\omega+\omega' = \Omega$. The dynamics of each frequency mode $\omega$ can thus be treated independently, and are given by  Equations  (\ref{eq:dndz}). 

A few other remarks about Equations  (\ref{eq:dndz}) are in order. 
Although the non-Hermitian parameter $\gamma_b$ doesn't appear in Equation (\ref{eq:dndz}a), it affects the dynamics of $n_a(\omega)$  through the coupling of the three different quantities in Equations (\ref{eq:dndz}). The set of coupled ODE's retains the quantum features expressed by the Lindblad master Equation (\ref{Lindblad}) through the last term in Equation (\ref{eq:dndz}c) that arises due to vacuum fluctuations and acts as a drive. Given the initial condition $n_a=n_b=m=0$, the term $-i\xi$ will force $m$ to acquire non-zero values, which in turn will drive $n_a$ and $n_b$ to finite values. This is in contrast to the classical counterpart of this processes, in which classical vacuum (i.e. zero fields) are steady state solutions of the nonlinear  problem (see Appendix \ref{sec:E}). 

We  can gain  intuition about the nature of the light generated by the above process by examining some properties of the equations of motion.  An important feature of  Equations (\ref{eq:dndz}), is that the second moments form a closed system, which indicates that the quantum operators exhibit linear dynamics (this is a consequence of treating the pump classically, and can be seen from the Heisenberg equations of motion for the operators $a$ and $b$). As a result, a system prepared in a Gaussian state (i.e. a state described fully by first- and second-order moments) \emph{remains} Gaussian.  

It is also interesting to consider the Heisenberg equations of motion for $\braket{a(\omega')}$ and  $\braket{b(\omega')}$.  In  Appendix \ref{sec:classPT}, we discuss how the equations of $\braket{a(\omega')}$ and  $\braket{b(\omega')}$ connect the current work to  recent activities in non-Hermitian physics. For our purposes here, we note that the equations of motion for $\braket{a(\omega')}$ and  $\braket{b(\omega')}$ do not have a noise term. Therefore, if the input state is a state with zero mean, the first-order moment remains  zero. Concretely, if the input state is a vacuum state, the generated light is necessarily described by a two-mode squeezed vacuum state, and thus the reduced state of each mode is a pseudothermal state---i.e. a state with an arbitrary spectrum (not necessarily blackbody), but with thermal photon-number statistics at each frequency \cite{Quesada2019}. 

Equations (\ref{eq:dndz}) also contain information about the correlations between the two generated beams. Non-zero $m$ implies that the two beams are correlated. To quantify to which degree these correlations are non-classical, we need  a  measure of quantum correlations such as the Entanglement of Formation (EoF) \cite{bennett1996mixed}. The EoF quantifies the entanglement of a state in terms of the entropy of entanglement \cite{janzing2009entropy} of the least entangled pure state needed to prepare it; it is zero for separable states, and increases with the amount of entanglement. There is no known analytical expression for the EoF for the states considered here, (analytical expressions for the EoF exist only for special cases). We thus compute it numerically using the approach recently introduced by Tserkis \emph{et al.} \cite{tserkis2019quantifying}.


\section{Results} 
In this section, we present the main results of this work, which are obtained by solving Equations (\ref{eq:dndz}) numerically. Importantly, we identify the regimes of operation where introducing loss in mode $b$ results in increased intensity of pseudothermal light generated in mode $a$.  We use the EoF to quantify the  entanglement between the beams, and show that they remain entangled, even after the addition of loss. We also compare our quantum results to known results for analogous classical systems.

First we note that when $\gamma_b=0$, Equations  (\ref{eq:dndz}) have a simple analytical solution:  $n_a(\omega)=\sinh^2(r(\omega))$ where $r(\omega)$ is a sinc function (in the low gain regime) whose width is inversely proportional to the length of the nonlinear medium, and also depends nontrivially on the medium's optical dispersion. To obtain solutions that apply to any material, we parameterize the spectral density as a function of $\Delta k$, which itself contains the medium's optical dispersion dependence. Finally, we note that  the spectral density is symmetric about the phase-matched point $\Delta k=0$, and so we only plot results for positive $\Delta k$.

\begin{figure}[t!]
\begin{center}
\includegraphics[width=\columnwidth]{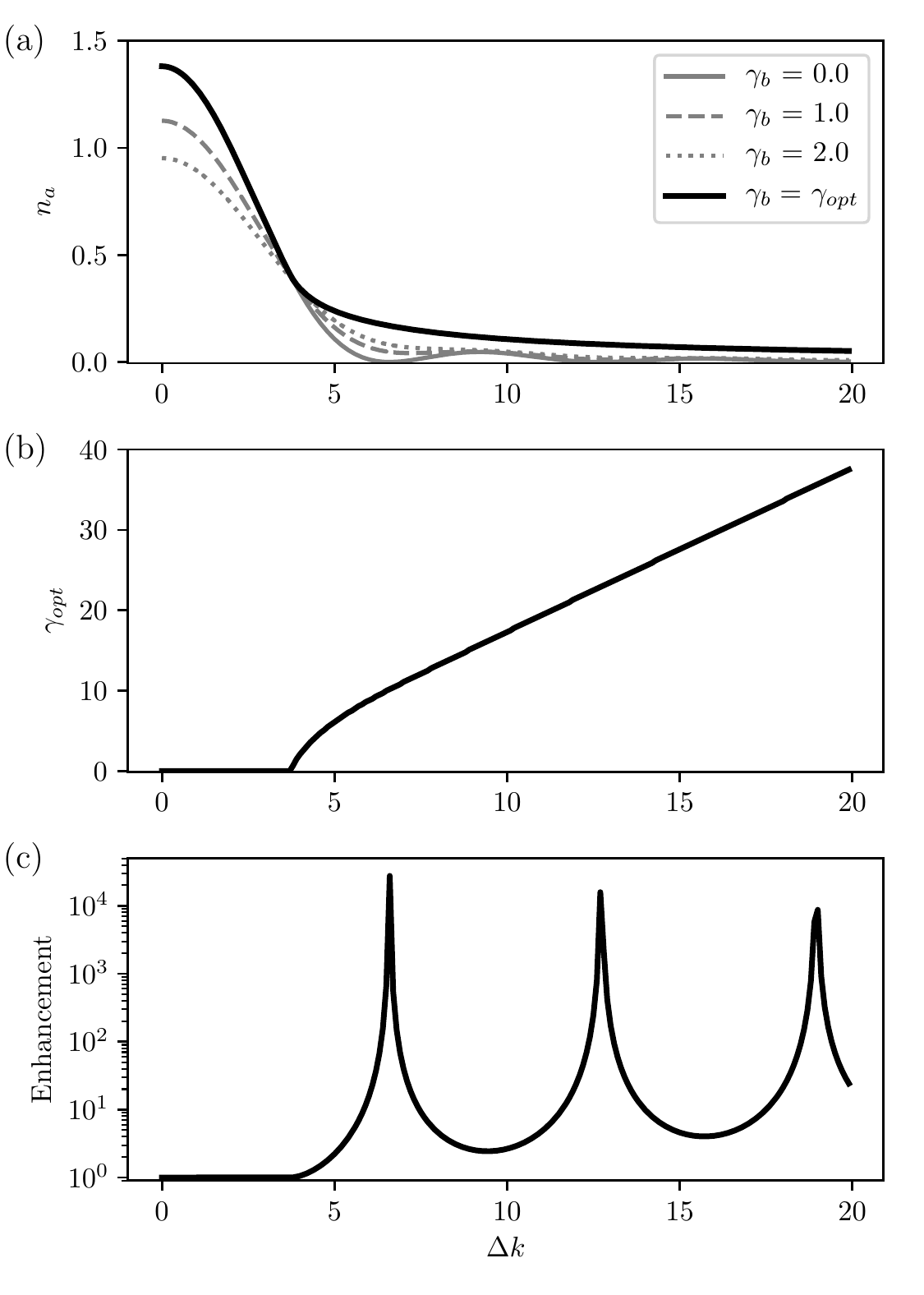}
\end{center}
\vspace{-0.3cm}
\caption{(a) Spectral density $n_a$ in mode $a$ for various loss rates $\gamma_b$ in mode $b$. (b) Optimal loss rate $\gamma_{\mathrm{opt}}$. Loss becomes beneficial away from phasematching ($\Delta k=0$). (c) Enhancement due to loss defined as $n_a$ for $\gamma=\gamma_{\mathrm{opt}}$ divided by $n_a$ for $\gamma=0$. Other parameters are $\chi=1.0$ and $\Delta k=11.5$. The enhancement factor defined by the output state value of $n_a^{\gamma_{opt}}/n_a^{\gamma=0}$ is shown in (c). Clearly, several orders-of-magnitude improvement can be observed outside the standard phase-matching regime.}
\label{fig:na_deltak}. 
\end{figure}

Figure \ref{fig:na_deltak} (a) plots the spectral density in mode $a$ as a function of $\Delta k$ for different loss rates $\gamma_b$. Within the phase-matching region (around $\Delta k=0$), the addition of loss does  not increase  the spectral density $n_a$ in mode $a$. In this regime, minimizing the loss will optimize the device performance. However, outside the phase-matching domain, introducing loss in mode $b$ can be beneficial, eventually leading to brighter light in mode $a$.  For the parameters used here (see figure captions), the  transition between these regimes occurs at $\Delta k\approx 4$. The black line shows the maximum achievable spectral density, which is obtained by using the optimal loss rates  $\gamma_{opt}$, plotted in Figure \ref{fig:na_deltak} (b). The enhancement factor defined by the output state value of $n_a^{\gamma_{opt}}/n_a^{\gamma=0}$ is shown in Figure  \ref{fig:na_deltak} (c). Clearly, several orders of magnitude improvement can be observed outside the standard phase-matching regime.
\begin{figure}[b!]
	\begin{center}
		\includegraphics[width=\columnwidth]{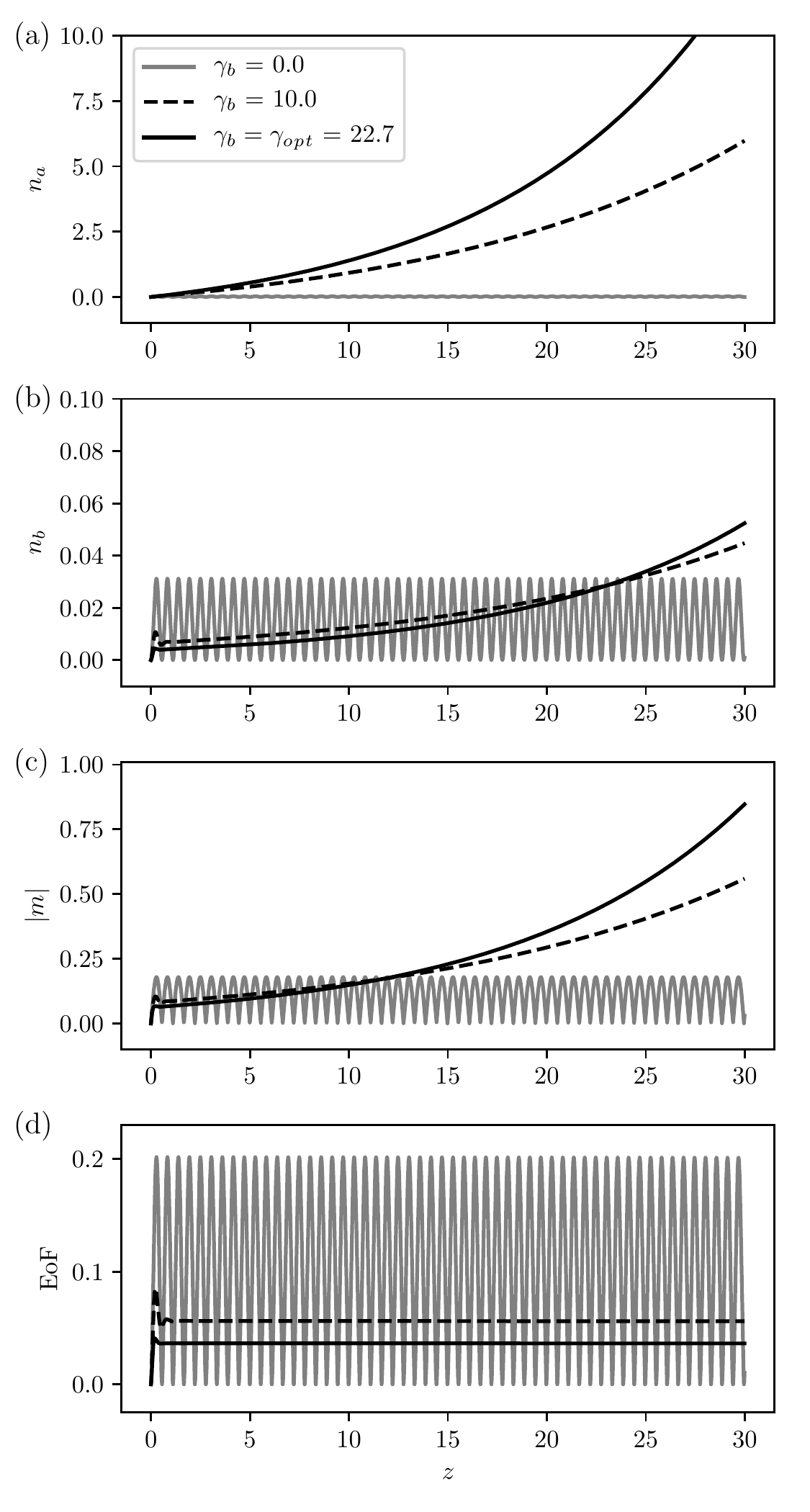}
	\end{center}
	\vspace{-0.3cm}
	\caption{Properties of the generated light for various values of loss parameter  $\gamma_b$. (a) Spectral density $n_a$ in mode $a$. (b) Spectral density $n_b$ in mode $b$. (c) The correlation parameter $m$. (d) The entanglement of formation (EoF) between modes $a$ and $b$. Other parameters are $\chi=1.0$ and $\Delta k=11.5$. Vertical axes have different scales, while horizontal axes are the same.}
	\label{fig:no_seed_comp}
\end{figure}

\begin{figure*}[t!]
	\begin{center}
		\includegraphics[width=\columnwidth]{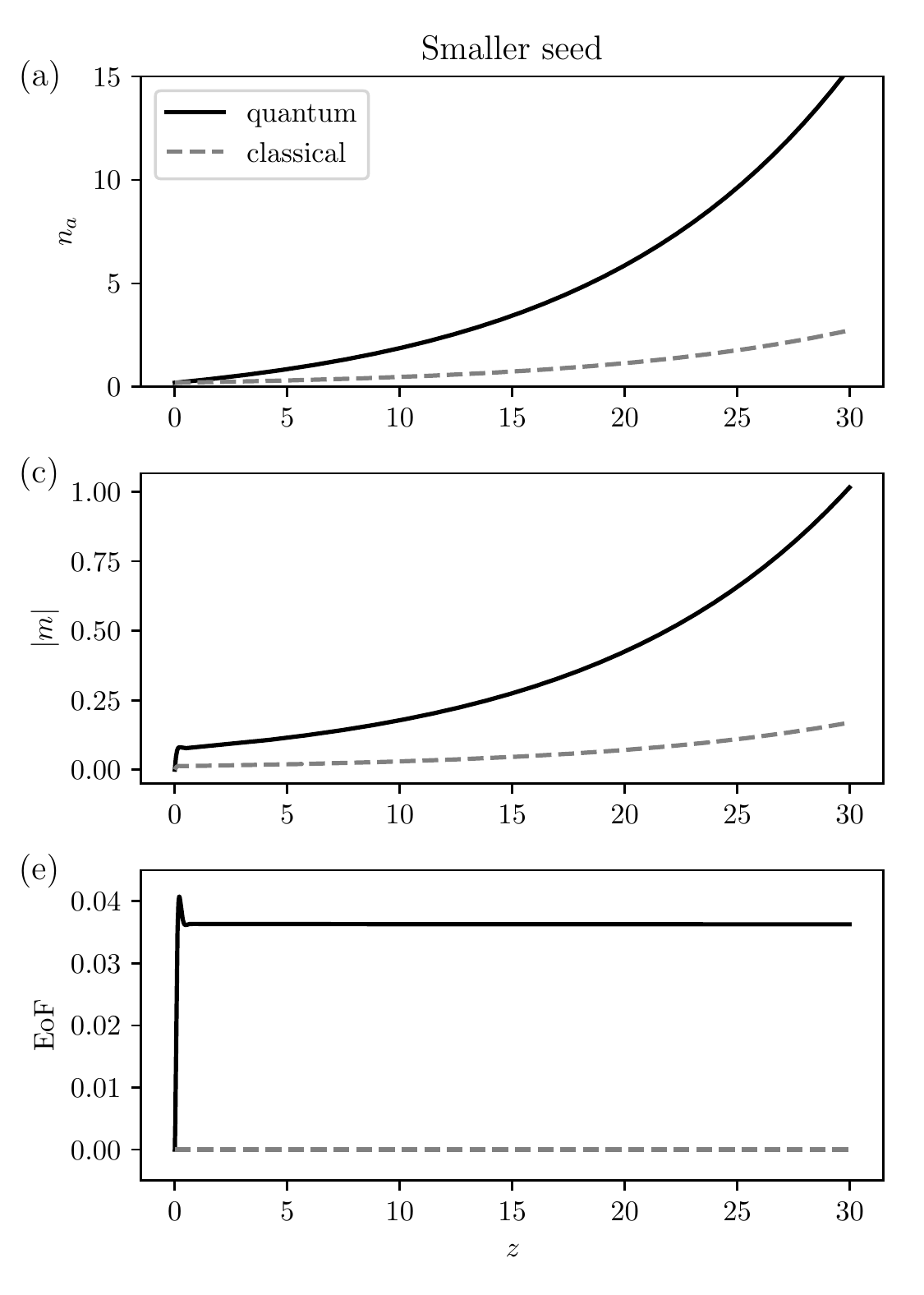}	\includegraphics[width=\columnwidth]{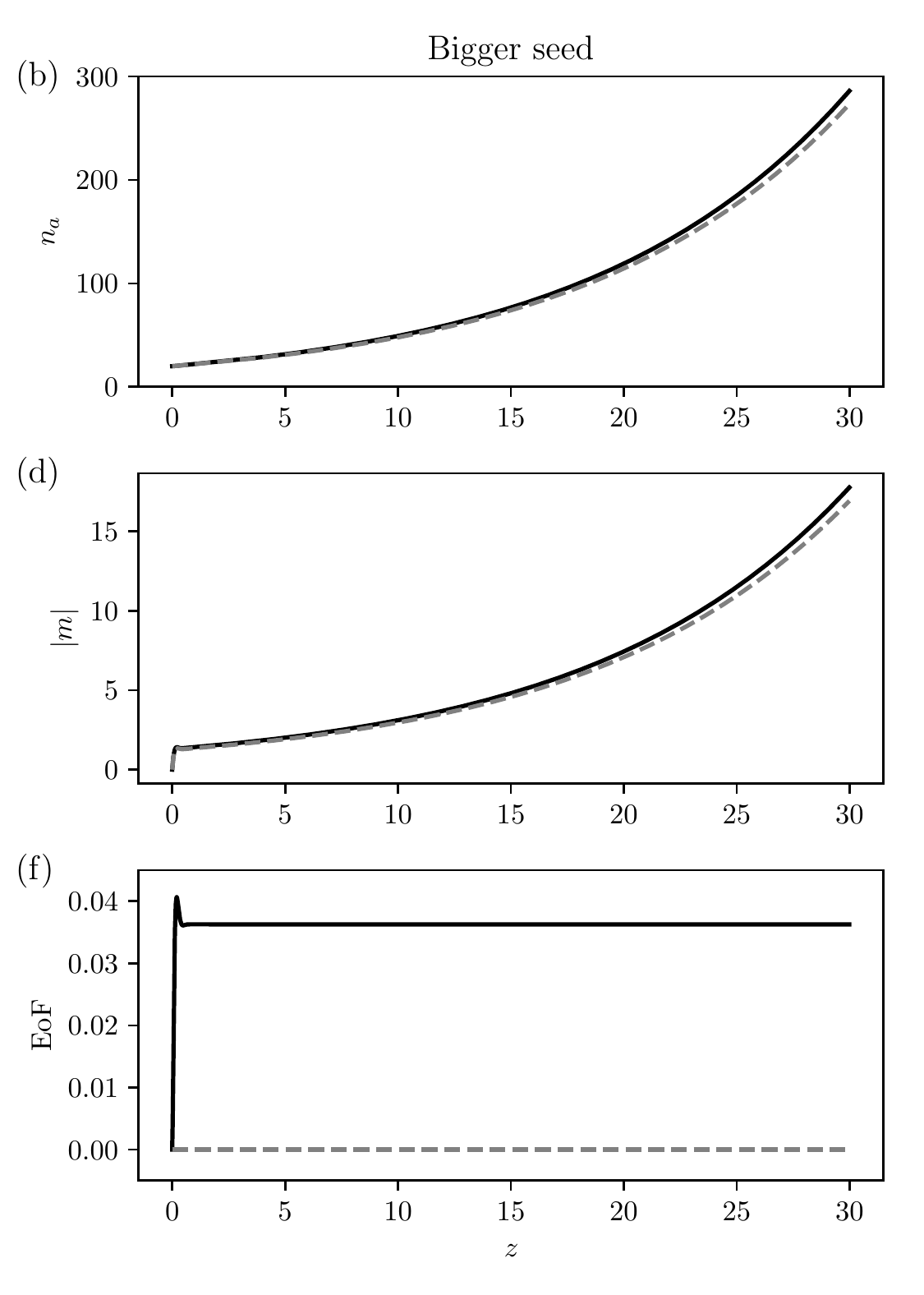}
	\end{center}
	\vspace{-0.3cm}
	\caption{Comparison between output predicted by quantum and classical models, using optimal loss parameter $\gamma=\gamma_{opt}=22.7$, squeezing parameter $\chi=1.0$ and phase mismatch $\Delta k=11.5$. 
	The spectral density $n_a$ in mode $a$  is plotted for smaller ($n_a(0)=0.2$) and bigger ($n_a(0)=20$) seeds in a) and b) respectively. The correlation parameter $m$ is plotted for smaller and bigger seeds in (c) and (d). The entanglement of formation (EoF) between modes $a$ and $b$ is plotted for smaller and bigger seeds in (e) and (f) respectively. For output predicted by the classical model, the EoF evaluates to be numerically zero, indicating no entanglement, as expected. Horizontal axes are the same in all plots.}
	\label{fig:seed}
\end{figure*}

Quite often, one is interested in one specific frequency. For a given material, this will correspond to a specific value of $\Delta k$. We therefore also investigate the dynamical features of the system for a given $\Delta k$. Figure \ref{fig:no_seed_comp} (a) shows that for a lossless system outside the phase-matching regime,  the spectral density in mode $a$ oscillates along the propagation direction $z$, remaining bounded. On the other hand, when loss is introduced in mode $b$, the spectral density in mode $a$ starts to grow, with the maximum amplification rate occurring at $\gamma_b=\gamma_{opt}$. Interestingly, for very large values of $\gamma_b$, beyond its optimal value, the signal amplification is suppressed (not shown). As we mentioned before, this occurs due to the competition between the pair generation and recombination rates. Informally, this behavior is similar to that of critical coupling in microcavities where the incident light completely couples to the cavity for an optimal value of the optical mode loss. Our analysis shows similar behaviour for $n_b$, as shown in Figure  \ref{fig:no_seed_comp} (b),  but on a much slower scale due to the direct effect of  loss on mode $b$. Figure \ref{fig:no_seed_comp} (c) shows that the correlation between the two modes also grows with the addition of loss. However, this does not reveal whether the two modes are entangled. To determine this, we plot the EoF. Interestingly, we find that it is always positive, indicating that the two modes are indeed entangled. 

Finally, we also compare the quantum system described in this paper with the analogous classical system. Classically, one typically solves equations of motion for the field amplitudes. Such equations can be recast into a form  similar to those in Equations (\ref{eq:dndz}); but with the $-i\xi$ term in Equation (\ref{eq:dndz}c) omitted. Up until now, we have only considered seedless initial conditions ($n_a(0)=n_b(0)=0$). For a classical system, these conditions predict no output in modes $a$ and $b$. To make the comparison, we therefore  consider a non-zero seed in mode $a$. Figure \ref{fig:seed} compares the quantum and classical cases for smaller ($n_a(0)=0.2$) and larger ($n_a(0)=20$) seeds. Figures  (\ref{fig:seed}a) and  (\ref{fig:seed}b) show that predictions for the spectral density $n_a$ in mode $a$ differ quite drastically between the quantum and classical models for a small seed, but converge for a larger seed. Figures  (\ref{fig:seed}c) and (\ref{fig:seed}d) show that predictions for the correlation between the modes also differ quite drastically between the quantum and classical models for a small seed, but converge for a bigger seed.  Figures (\ref{fig:seed}e) and  (\ref{fig:seed}f), however, show that when considering entanglement between the modes, predictions of the two models differ both for smaller and bigger seeds; the output predicted by classical models does not contain entanglement, as should be expected, while the output predicted by the quantum model does. Interestingly, the amount of entanglement seems independent of the size of the seed.  This also shows that the introduction of loss does not completely destroy entanglement between the modes. 

\section{Conclusion} 

Most prior studies of non-Hermitian engineering in quantum optical systems emphasized the limitations imposed by  quantum  noise  on  non-Hermitian  systems. In this paper, we asked if there exist situations where optical losses can be useful in quantum-state engineering. 

We addressed this question in the context of spontaneous generation of broadband pseudothermal  states  in  one  output  mode  of  a  wave-mixing  process, and showed that even when quantum fluctuations are present, non-Hermiticity can still play a positive role. Specifically, we showed that careful engineering of optical loss can be used to increase the brightness of broadband pseudothermal states, even in the absence of phase matching. We also showed that entanglement is present in the generated light, even in the presence of optical loss, distinguishing the process from optical-loss-induced amplification  in classical  systems. 

This work could be extended in a number of ways.  One could  consider the effect of optical loss on pseudothermal states with interesting coherence properties, such as those generated by pulsed pump lasers \cite{Quesada2019}. Furthermore, it would be interesting to consider frequency-dependent loss in the context of spectral shaping. We leave these for future research.

We expect our results to have applications in quantum-state generation for quantum technologies. While satisfying phase matching is in general favorable, it is not always possible. Our work opens the door for utilizing the full potential of semiconductor materials (such as silicon and AlGaAs) that exhibit giant nonlinearities but lack the necessary ingredients for achieving quasi-phase matching (see Appendix \ref{sec:F} for more discussion on possible implementations). This in turn may pave the way for using these platforms to build on-chip quantum light sources.   

\section{Acknowledgements}
AMB thanks Ish Dhand for interesting conversations. We also thank Spyros Tserkis for answering questions about the numerical simulations of the EoF. Research at Perimeter Institute is supported by the Government of  Canada through Industry Canada and by the Province of Ontario through the Ministry of Research and Innovation. We acknowledge the support of the Natural Sciences and Engineering Research Council of Canada.


\appendix
\onecolumngrid

\section{Motivation for the nonlinear momentum operator}\label{sec:H}
In this section, we derive the equation of motion for photon-number expectation values of the two output modes $a$ and $b$. 

Our starting point is the effective momentum operator that generates the $z$ dynamics of the operators $a(z,\omega)$ derived in \cite{Quesada2018b}. The operators $a(z,\omega)$ satisfy the commutation relations $[a_k(z,\omega),a_{k'}\dg(z,\omega')]=\delta_{k,k'}\delta(\omega-\omega')$, and  can be thought of as field operators that annihilate photons at frequency $\omega$. 

The expression in \cite{Quesada2018b} includes  cross-phase modulation. We do not take this into account. The effective momentum operator for our system is:
\begin{align}
P ={}& \hbar \int d\omega [\Delta k_a (\omega) a\dg(\omega) a(\omega)+\Delta k_b (\omega) b\dg(\omega) b(\omega)]  + \hbar \int \int d\omega d\omega' f(z,\omega,\omega') a\dg(\omega)b\dg(\omega')+\text{H.c.} \,,
\end{align}
where $f(z,\omega,\omega')$ is a function that depends on the nonlinearity of the material, the pump spectral amplitude, and the group velocities of the fields. The functions $\Delta k_j(\omega)$ determine the phase matching inside the material:
\begin{align}
\Delta k_j(\omega) = \left( \frac{1}{v_j} -\frac{1}{v_p} \right)(\omega-\bar{\omega}_j), \quad \text{for } j={a,b}
\end{align}
and for the fields $a,b$ and the pump we have written their dispersion relation as
\begin{align}
k - \bar{k}_\mu  = \frac{\omega -\bar{\omega}_\mu}{v_\mu} \quad \text{for } \mu = {a,b,p}
\end{align}
 where we have neglected group velocity dispersion within each field $a,b,p$ and we wrote the group velocity of each field as $v_\mu$. 
 Furthermore we assume that the central wavevectors and frequencies of the three fields participating in the nonlinear process satisfy
 \begin{align}
\bar{\omega}_a+\bar{\omega}_b-\bar{\omega}_p=0, \quad 
\bar{k}_a+\bar{k}_b-\bar{k}_p=0
 \end{align}
for SPDC or 
  \begin{align}
  \bar{\omega}_a+\bar{\omega}_b-2\bar{\omega}_p=0, \quad 
  \bar{k}_a+\bar{k}_b-2\bar{k}_p=0
  \end{align}
 for SFWM.
 
Since we are considering a cw pump, we take $f(z,\omega,\omega')=\xi g(z)\delta(\omega+\omega'-\Omega)$, where $\xi$ (which depends on the nonlinearity strength and the amplitude of the pump) determines the strength of the interaction, and $g(z)$ is the normalized nonlinearity profile of the material and 
here, we take it to be a rectangular function $\Pi_{0,L}(z)=1$ for $0<z<L$ and $\Pi_{0,L}(z)=0$ otherwise.
Finally, 
\begin{align}
\Omega = \begin{cases}
2 \bar{\omega}_p = \bar{\omega}_a + \bar{\omega}_b \text{ for SFWM,}\\
\bar{\omega}_p = \bar{\omega}_a + \bar{\omega}_b \text{ for SPDC.}
\end{cases}
\end{align}
This gives:

\begin{align} \label{P_appendix}
P    ={}& \hbar \int d\omega [\Delta k_a (\omega) a\dg(\omega) a(\omega)+\Delta k_b (\omega) b\dg(\omega) b(\omega)]  + \hbar \xi g(z) \int d\omega a\dg(\omega)b\dg(\Omega-\omega)+\text{h.c.} 
\end{align}

\section{Broadband Pseudothermal states}\label{BPT}

Nonlinear processes such as spontaneous parametric down conversion (SPDC) or spontaneous four wave mixing (SFWM) can generate spectrally entangled twin beams.
The reduced state of each beam, obtained by tracing out the other component, can be thought of as a broadband ``pseudothermal'' state \cite{Quesada2019} whose spectral coherence can be tuned---from perfect coherence to complete incoherence---by adjusting the pump spectral width. 

In the limit of a cw laser pumping the twin beam's quantum state is
\begin{align}
\ket{\psi} = {}&\mathcal{\hat U}_{\text{SQ}} \ket{\text{vac}}\,;\\
\mathcal{\hat U}_{\text{SQ}} ={}  & e^{ \left(\int \mathrm d \omega\, r(\omega) \hat{a}^\dagger(\omega) \hat{b}^\dagger(\Omega -\omega) - \hc \right) }\,,
\end{align}
where $r(\omega)$ is the dispersive nonlinear coupling coefficient and it is a function of optical properties of the material \cite{Quesada2019} (in the low-gain regime, $r(\omega)$ is  the phase-matching function parameterized in terms of $\omega$). The reduced state of, say, beam $a$ is then given by:
\begin{align}
\rho_a &= \frac{1}{Z}e^{- \int \mathrm d \omega \alpha(\omega) \hat a^\dagger (\omega) \hat a(\omega) }\,,\\
Z&=\text{Tr}\left(e^{- \int \mathrm d \omega \alpha(\omega) \hat a^\dagger (\omega) \hat a(\omega) }\right)\,,
\end{align}
where $\alpha(\omega) = \log \left( 1/ \tanh^{2}(r (\omega)) \right)$. Note that for $\alpha(\omega)=\hbar\omega/k_B T$, $\rho_a $ represents a multi-mode thermal state in the traditional sense.

The spectral density of the pseudothermal state is: 
\begin{align}\label{eq:ncw}
n(\omega)& =\braket{\hat a^\dagger(\omega) \hat a(\omega')}_\psi= \sinh^2(r(\omega))\,.
\end{align}
In general, this is a  peaked function that becomes higher and narrower as the length $L$ of the nonlinear region increases. For $\omega$ ranging under the peak value, the system is phase-matched, and the intensity of the thermal state grows with $L$. however, outside this favorable operation bandwidth, destructive interferences between wave components impedes this growth by providing  scattering channels for the reverse process.

Traditionally, this problem is often addressed by  engineering $r(\omega)$ using quasi-phase-matching. This technique however does not lend itself to easy implementation in semiconductor platforms that do not exhibit electric domains. Given the giant nonlinear coefficients of these material platforms, and the potential future for silicon photonics and hybrid integration, it would be of immense interest to device a different route around this obstacle. In the main text, we show how loss engineering can come to aid.

\section{Derivation of the equations of motion}\label{app:eoms}

Starting from the following form of the master equation 
\eq{
\frac{d}{dz}\langle \mathcal{O} \rangle ={}& -\frac{i}{\hbar} \langle [\mathcal{O}, P] \rangle -\sum_{c \in \{a,b \}} \frac{\gamma_c}{2} \int d\omega''  \Big\langle c^\dagger(\omega'') [c(\omega''), \mathcal{O}] + [\mathcal{O}, c^\dagger(\omega'')] c(\omega'')\Big\rangle,
}
with $P$ defined in Equation (\ref{P_appendix}).

We investigate the dynamics of the expectation values
\seq{\eq{\label{moments}
&\langle a^\dagger(\omega) a(\omega')\rangle\\ &\langle b^\dagger(\Omega-\omega) b(\Omega-\omega')\rangle\\ &\langle a(\omega) b(\Omega-\omega')\rangle,
}}
for which we find
\seq{\label{eoms}
\eq{
\begin{split}
\frac{d}{dz} \braket{a^\dagger(\omega) a(\omega')} =& -i \xi g (z) \braket{a^{\dagger }(\omega )b^{\dagger }\left(\Omega -\omega'\right)}+i \xi g (z) \braket{a\left(\omega' \right)b(\Omega -\omega )}\\
&+i \left\{  \Delta k _{a}(\omega ) - \Delta k _{a}\left(\omega'\right)\right\} \braket{a^{\dagger }(\omega )a\left(\omega'\right)} -\gamma_a 	\braket{a^\dagger(\omega) a(\omega')},
\end{split}\\
\begin{split}
\frac{d}{dz} \braket{b^\dagger(\Omega-\omega)b(\Omega-\omega')}  =& 	-i \xi g (z) \braket{ a^{\dagger }\left(\omega'\right)b^{\dagger }(\Omega -\omega )}+i \xi g (z) \braket{a(\omega)b\left(\Omega -\omega'\right)}\\
&+i \left\{ \Delta k _{b}(\Omega -\omega ) -\Delta k _{b}\left(\Omega -\omega'\right) \right\}\braket{b^{\dagger } (\Omega -\omega)b\left(\Omega -\omega'\right)}\\
&-\gamma_b 	\braket{b^\dagger(\Omega-\omega) b(\Omega-\omega')},
\end{split}\\
\begin{split}
\frac{d}{dz} \braket{a(\omega) b(\Omega-\omega')} =&-i \xi g (z)
\braket{a^{\dagger }\left(\omega'\right)a(\omega )}
-i \xi g (z) \braket{ b^{\dagger }(\Omega -\omega )b\left(\Omega -\omega'\right)}\\
&-i \left\{\Delta k_{a}(\omega ) +i \Delta k_{b} \left(\Omega -\omega'\right)\right\} \braket{a(\omega ) b\left(\Omega -\omega'\right)}\\
&-\frac{\gamma_a+\gamma_b}{2} \braket{a(\omega) b(\Omega-\omega')}-i \delta\left(\omega -\omega'\right) \xi g (z).
\end{split}
}
}
Note that the only inhomogeneous term is $-i \delta\left(\omega -\omega'\right) \xi g (z)$. If it was not for this term, that drives vacuum fluctuations, the correlation functions in Equation (\ref{moments}) would remain zero for all time if they are zero at time $t=0$. This inhomogeneous term also tells us that the only ``slice'' of the correlation functions that is driven to a nonzero value is the one for which $\omega=\omega'$. Based on the preceding argument, we introduce the following notation:
\seq{\eq{
\braket{a^\dagger(\omega) a(\omega')} = \delta(\omega-\omega') n_a(\omega),\\
\braket{b^\dagger(\Omega-\omega) b(\Omega-\omega')}=\delta(\omega-\omega') n_b(\omega),\\
\braket{a(\omega) b(\Omega-\omega')}=  \delta(\omega-\omega') m(\omega).
}}
Inserting these expressions into Equations   (\ref{eoms}) we obtain  the equations of motion  in Equations (\ref{eq:dndz}).

\section{Classical limit}\label{sec:E}
For completeness, we briefly discuss the classical limit of quantum process under investigation. Within the undepleted pump approximations (which allows us to neglect variations in the pump component), and small signal limit (which allows us to neglect the self- and cross-phase modulations for the signal and idler beams), the classical equations for the signal and idler beams are
\begin{align}\label{N}
\frac{d}{dz} \mathbf{u} &= i \mathbf{N} \mathbf{u};~~\mathbf{N}=\left(
\begin{array}{cc}
\frac{i \gamma _a}{2}-\Delta k_a & \zeta 
\\
-\zeta  & \frac{i \gamma _b}{2}+\Delta k_b
\\
\end{array}
\right)\,, 
\end{align}
where $\mathbf{u} = (\alpha,\beta^*)^T$, and where $\alpha$ and $\beta$ are the field amplitudes.

If we multiply the first row by $\alpha^*$ and add the resultant equation to its complex conjugate, we recover Equation (\ref{eq:dndz}a). Similarly we can recover Equation (\ref{eq:dndz}b) by multiplying the second row by $\beta^*$ and adding the resultant equation to its complex conjugate. On the other hand, multiplying the first/second row by $\beta$ and $\alpha$, respectively and adding the results, gives Equation (\ref{eq:dndz}c) but without the drive term arising from the quantum noise (as one would expect).

Put differently, the classical limit can be obtained from the quantum description by using the factorizations $\braket{a\dg(\omega)a(\omega')}=\braket{a\dg(\omega)} \braket{a(\omega')}$, $\braket{b\dg(\omega)b(\omega')}=\braket{b\dg(\omega)} \braket{b(\omega')}$, $\braket{a(\omega)b(\omega')}=\braket{a(\omega)} \braket{b(\omega')}$; and neglecting the noise term.

\section{Field equations and relation to PT symmetry}\label{sec:classPT}

By using a simple gauge transformation, the matrix $\mathbf{N}$ in Equation (\ref{N}) can be cast in a more useful form:
\begin{align} \label{N_Prime}
\mathbf{N}'&=\left(
\begin{array}{cc}
\eta & \zeta 
\\
-\zeta  & -\eta
\\
\end{array}
\right)\,,
\end{align}
with $\eta=(i\Delta \gamma - \Delta k)/2$, where $\Delta \gamma=(\gamma_a-\gamma_b)/2$ and $\Delta k=\Delta k_a + \Delta k_b$. When $\gamma=0$, the system in (\ref{N_Prime}) exhibits an exceptional point at $\eta=\zeta$, which marks the transition between the phase-matching regime ($\eta \le \zeta$) and the phase mismatch domain. In the former, the signal experience amplification while in the latter, dynamics are oscillatory. This behavior, which can be emulated by a linear wave-guide array \cite{Sipe2013PRA} is not accidental. In fact, while $N'$ (with $\Delta \gamma=0$) does not respect parity-time reversal (PT) symmetry (reflecting its SU(1,1) symmetry as opposed to SU(2) in the case of PT systems), it satisfies a generalized PT condition \cite{Miri2016}. Particularly, when $\Delta \gamma=0$, $[PT,SN'S^{-1}]=0$  where the coefficients of the matrix S are given by $S_{11}=S_{21}=1$ and $S_{12}=-S_{22}=-i$. 

By introducing a finite value for $\gamma$, the behavior of the eigenvalues are different, always exhibiting an imaginary component that leads to signal amplifications, as studied in detail in \cite{El-Ganainy2015OL,Zhong2016NJP}. 

A final interesting remark on Equation (\ref{N}) is that it is exactly identical to the classical description of undepleted pump 4-wave mixing in the small signal limit. In other words, in contrast to (\ref{eq:dndz}), it does not contain any noise term that signifies the quantum origin of the problem. Thus, the generated light will have intensity but the quantum expectation values of the field operators will always remain zero--- a characteristic feature of thermal states.

\section{Practical considerations} \label{sec:F}

Our discussion in the main text has focused on an ideal situation where the pump and signal losses are absent. In any real system, the optical loss will not be completely absent at any given frequency. However, a judicious choice of the material system along with the geometric parameters can render the absorption at these two frequencies negligible withing the specific propagation distance. For example, a $5\%$ variation of the pump power across the propagation direction will not have a significant impact on the conversion efficiency. More over, the optical loss at the signal frequency must be significantly smaller than the nonlinear gain. From a practical perspective, these two conditions can be satisfied easily and indeed they are met in most experimental work that include generation of entangled photon pairs. 

What is more difficult is to satisfy these two conditions along with the required optical loss of the idler component. In this respect, several implementation strategies can be attempted. First, one may rely completely on the material properties. For instance, early experimental works on classical wave-mixing under imperfect phase matching conditions showed that it is possible to increase the output signal power by using standard silica fiber due to the higher idler loss \cite{jauregui2012high}. Another possibility is to synthesize these absorption features artificially. In previous studies \cite{El-Ganainy2015OL,Zhong2016NJP} related to the classical analogue of the effect considered here, we proposed an implementation based on two coupled waveguides that are mode matched only at the idler frequency. In this case, the idler component will experience a much stronger coupling from the main waveguide to the auxiliary one. By doing so, one can engineer the idler loss by either depositing metal films or imprinting radiation Bragg gratings on top of the auxiliary waveguide. Alternatively, one may even use only one waveguide with a radiation Bragg gratings that exhibit very narrow bandwidth centered at the idler frequency. Yet a third possibility is to use hybrid dielectric-plasmonic modes to introduce losses to the certain selected modes. We explore some of these possibilities in future work.  

\end{document}